\documentclass[
reprint,
amsmath,
amssymb,
aps,superscriptaddress
]{revtex4-2}

\usepackage{graphicx}
\usepackage{bm}
\usepackage{amsfonts}
\usepackage{hyperref}   
\usepackage{cleveref}   
\usepackage[normalem]{ulem}
\usepackage{xcolor}
\usepackage{footmisc}

\usepackage{amsmath,amssymb}
\usepackage{siunitx}
\usepackage{braket}
\usepackage{wrapfig}

\renewcommand\a{\alpha}
\renewcommand\b{\beta}
\newcommand\g{\gamma}
\renewcommand\d{\delta}

\renewcommand\i{\iota}
\renewcommand\k{\kappa}
\renewcommand\l{\lambda}

\newcommand\s{\sigma}
\renewcommand\t{\tau}
\renewcommand\u{\upsilon}

\renewcommand\c{\chi}
\renewcommand\j{\psi}
\renewcommand\o{\omega}
\newcommand{\dint}{{\rm d}}
\newcommand{\xpom}{x_{I\!\!P}}

\renewcommand\L{\Lambda}

\newcommand{\lan}{\langle}
\newcommand{\ran}{\rangle}

\newcommand{\diracslash}[1]{#1\llap{/\kern2pt}}

\newcommand{\be}{\begin{equation}}
\newcommand{\ee}{\end{equation}}
\newcommand{\bea}{\begin{eqnarray}}
\newcommand{\eea}{\end{eqnarray}}
\newcommand{\ba}[1]{\begin{array}{#1}}
\newcommand{\ea}{\end{array}}
\newcommand{\bep}{\begin{pmatrix}}
\newcommand{\eep}{\end{pmatrix}}

\newcommand{\bt}{\begin{tabular}}
\newcommand{\et}{\end{tabular}}

\newcommand{\beas}{\begin{eqnarray*}}
\newcommand{\eeas}{\end{eqnarray*}}

\begin{document}

\title{Imprints of nuclear shell structure in exclusive vector meson production}

\author{Arpita Mondal}%
\altaffiliation{These authors contributed equally to this work.}
\affiliation{Department of Physics, Indian Institute of Technology Bombay, Mumbai-400076, India}
\email{arpita.mondal@iitb.ac.in}

\author{Arjun Kumar}%
\altaffiliation{These authors contributed equally to this work.}
\affiliation{Center for Frontiers in Nuclear Science, Department of Physics and Astronomy, Stony Brook University, New York 11794--3800, USA}
\email{arjun.kumar.2@stonybrook.edu}

\author{Debojit Sarkar}
\email{debojit.sarkar@iitb.ac.in}
\affiliation{Department of Physics, Indian Institute of Technology Bombay, Mumbai-400076, India}


    	\date{\today}

    	\begin{abstract}

We report for the first time that, within the saturation framework, exclusive vector meson production at small $x$ is sensitive to the shell structure of the target nucleus. 
Self-consistent nuclear densities from occupied single-particle orbitals in the quark-meson coupling (QMC) model modify the coherent $|t|$-differential cross section, enhancing secondary diffractive lobes in light nuclei and displacing the higher-order minima in intermediate-mass nuclei, whereas for heavy targets the shifts are weaker.
Because the small $J/\psi$ dipole is insensitive to saturation, these features make coherent $J/\psi$ production a clean probe of nuclear shell structure, most favorably for intermediate-mass nuclei such as calcium isotopes.
For the larger $\phi$ dipole, shell structure must be included in the nuclear initial state before saturation effects can be isolated in differential observables at the Electron Ion Collider.
Our results establish exclusive vector meson production as a probe of the mean-field nuclear structure at small $x$ and provide a baseline for isolating residual many-body correlations.

\end{abstract}

\maketitle

\paragraph*{Introduction.}Understanding the nuclear structure and how different nuclear interactions give rise to the geometric structure of nuclei is a central challenge at the intersection of nuclear physics and quantum chromodynamics (QCD). On one hand, heavy ion collisions at the Relativistic Heavy Ion Collider (RHIC) and the Large Hadron Collider (LHC) constrain the geometry indirectly through final-state multi particle correlations \cite{FSgeom2,FSgeom3,FSgeom4,FSgeom1,Schenke:2012hg,Schenke:2012wb}, whereas the nuclear deep inelastic scattering (DIS) program provides a complementary and more direct probe of the same initial geometry. The forthcoming Electron-Ion Collider (EIC) \cite{Accardi:2012qut,AbdulKhalek:2021gbh} in the United States will realize high-energy $e$A collisions in collider kinematics, thereby enabling a controlled extraction of nuclear geometry for the first time in this regime.

In the so-called diffractive events, owing to the large coherence length of the virtual photon at high energies, the projectile wave function interacts with the nucleus as a whole, leaving the target intact. At the same time, the underlying interaction proceeds via two-gluon or multiple two-gluon exchange, making it highly sensitive to non-linear effects. Exclusive vector meson production \cite{Aktas:2005xu,Chekanov:2002xi,ZEUS:2007iet} constitutes a particularly clean subset of this class of events, with a three-body final state consisting of the scattered lepton, a vector meson and the intact nucleus. The differential $t$-spectrum in such events exhibits characteristic diffractive minima whose positions encode both nuclear geometry and nonlinear QCD dynamics. Precision measurements of these minima and the extraction of the associated nuclear thickness constitute one of the golden channels at the EIC \cite{Toll:2012mb,Toll:2013gda}.

The atomic nucleus is a correlated quantum many-body system whose ground-state wave function contains, simultaneously, single-particle shell structure, two and higher-body correlations, including short-range correlated pairs and cluster-like configurations. Different observables project onto different sectors and length scales of this wave function, making it essential to disentangle the one-body geometric structure from genuine many-body correlations. Recent electron-scattering measurements demonstrate that these scales are not independent, short-range nucleon pairing is strongly constrained by the occupied nuclear orbitals, establishing a direct connection between long-range shell structure and short-distance dynamics \cite{Nguyen2026NuclearShellStructure, Nguyen:2026ztp}. This observation motivates a corresponding question for exclusive vector-meson production at the EIC, to what extent is the differential $t$-spectrum governed by the shell structure encoded in the nuclear one-body density, and what part remains sensitive to residual many-body correlations?  

Earlier investigations have shown that the diffractive $t$-spectrum is a sensitive probe for multiscale imaging of nuclear deformation across distinct transverse length scales \cite{multiscale}. Recent studies of light nuclei using \textit{ab initio} nucleon configurations have demonstrated shifts in the diffractive minima of exclusive vector-meson production within the Color Glass Condensate (CGC) framework \cite{abintio,lightnuclei}. However, although these configurations encode microscopic nuclear structure, they do not permit a straightforward interpretation in terms of individual occupied single-particle orbitals, complicating a direct connection between diffractive observables and nuclear shell structure. Moreover, the computational cost of \textit{ab initio} approaches limits their systematic application across the nuclear chart, particularly for heavy nuclei, making it challenging to investigate the evolution of shell-structure signatures across a broad range of nuclei. 

In this Letter, we employ the quark–meson coupling (QMC) model \cite{QMC1,QMC2,Guichon:1987jp, Saito:2005rv}, a QCD-motivated mean-field framework to compute nuclear densities self-consistently from occupied single-particle orbitals, thereby providing a direct connection to shell structure across a broad range of nuclei. Using QMC densities for $^{16}\mathrm{O}$, $^{40}\mathrm{Ca}$, $^{48}\mathrm{Ca}$, and $^{208}\mathrm{Pb}$, we calculate coherent exclusive $J/\psi$ and $\phi$ production in $eA$ collisions using the color-dipole framework~\cite{Nikolaev:1990ja,Mueller:1993rr} which provides a unified framework for inclusive and diffractive scattering at high energies incorporating unitarity. The primary aim of this work is to quantify how much of the diffractive structure in exclusive vector-meson production is already generated by nuclear shell filling, prior to correlations and many-body dynamics. We present model predictions for the $t$-spectrum relevant to future EIC measurements and propose to measure $t$-spectrum of nuclear isotopes to probe the shell structure.\\

\paragraph*{Exclusive diffraction off nucleus at high energy.} In exclusive diffraction, the large rapidity gap between the produced meson and the scattered nucleus reflects the color-singlet exchange in the $t$-channel, mediated at leading order by at least two gluons. In the dipole picture, the virtual photon fluctuates into a ($q\bar q$) dipole, which scatters elastically from the target and subsequently projects onto the vector-meson wave function. The scattering amplitude is Fourier conjugate to the transverse momentum transfer $(\boldsymbol{\Delta})$, with ($t=-\Delta^{2}$). Consequently, the $t$-dependence of the differential cross section, (d$\sigma$/dt), provides direct sensitivity to the geometry of the target. The scattering amplitude is \cite{Kowalski:2003hm,Lappi:2010dd, Kumar:2021zbn,Kumar:2022aly}:
\begin{align}
 \label{amp}
    	\mathcal{A}^{\gamma^* A} (\xpom,Q^2,\textbf{$\Delta$})&=i\int \dint^2 \textbf{r}\int \dint^2\textbf{b}\int \frac{\dint z}{4 \pi} ~  e^{-i[\textbf{b}-(\frac{1}{2}-z)\textbf{r}].\textbf{$\Delta$}}  \\ \nonumber 
	&\times [\Psi^*_V\Psi](Q^2,\textbf{r},z) ~\frac{\dint\sigma _{q\bar{q}}^{(\Omega_i)}}{\dint^2\textbf{b}}(\textbf{b},\textbf{r},\xpom)
 \end{align}
where $\mathcal{A}$ is the scattering amplitude for the diffractive vector meson photo-production and the coherent cross section is given by the first moment of the amplitude:
\begin{equation}
\label{diff_cross}
    	\frac{\dint \sigma^{\gamma^* A }}{\dint t} = \frac{1}{16 \pi} \big| \lan \mathcal{A}^{\gamma^* A } \ran_{\Omega}\big|^2
\end{equation}
and $\lan ...\ran_{\Omega}$ denotes the average over nucleon configurations $\Omega_i$ of the target taken at the amplitude level. Here $z$ is the fraction of photon energy carried by the quark, $\xpom$ is the fraction of the target's longitudinal momentum transferred to the produced meson, \textbf{b} is the impact parameter of the dipole relative to the target center, $Q^2$ is the photon virtuality. The overlap $\Psi^*_V \Psi$ between the virtual-photon and vector-meson is computed using the Boosted Gaussian ansatz~\cite{Kowalski:2006hc}. In the IPSat model the dipole amplitude is given as \cite{Kowalski:2003hm}:
 \begin{eqnarray}
    	\frac{\dint\sigma _{q\bar{q}}^{(\Omega_i)}}{\dint^2\textbf{b}}=2\bigg[1-\text{exp}\big(-\frac{\pi^2}{2N_C} \textbf{r}^2 \alpha_s(\mu^2) \xpom g(\xpom,\mu^2)	T_A^{\Omega_i}(\textbf{b})\bigg]
\end{eqnarray}
with $T_A(\textbf{b},\Omega_i) = \sum_{i=1}^{A} T_p(\textbf{b}-\textbf{b}_i)$, the dipole amplitude saturates for large dipole size, large thickness, or high target gluon density. The strong coupling $\alpha_s$ and the gluon density \(\xpom g(\xpom,\mu^{2})\) are evaluated on the scale $\mu^2 = \mu_0^2 +\frac{C}{r^2}$, with the gluon density fit from \cite{Kowalski:2006hc}. For the first moment of the amplitude we use the analytical average of the dipole cross section \cite{Kowalski:2003hm,Toll:2012mb}
 \begin{eqnarray}
    	\bigg<\frac{\dint\sigma _{q\bar{q}}^{}}{\dint^2\textbf{b}}\bigg>_{\Omega}=
	2\bigg[1-\bigg(1-\frac{T_A( \textbf{b})}{2} \sigma_{q \bar{q}}^p \bigg)^A\bigg]
\end{eqnarray}
In the linearized (IPNonSat) limit, used to quantify non-linear effects, the analytical average is given as \cite{Toll:2012mb},
\begin{eqnarray}
    \bigg<\frac{\dint\sigma _{q\bar{q}}^{}}{\dint^2\textbf{b}}\bigg>_{\Omega} = 
 \frac{\pi^2}{N_C} \textbf{r}^2 \alpha_s(\mu^2) \xpom g(\xpom,\mu^2)AT_A(\textbf{b})
\end{eqnarray}
where the thickness $T_A(\textbf{b})$ is evaluated by projecting the density obtained from the QMC model in the transverse plane 
\begin{eqnarray}
    T_A(\textbf{b}) = \int_{-\infty}^{\infty} dz\,\rho(\textbf{b},z)~,~~ r = \sqrt{\textbf{b}^2 + z^2}
\end{eqnarray}

\paragraph*{Shell-structure densities.} 
In QMC, non-overlapping MIT bags are bound by self-consistent scalar and vector mean fields that couple directly to the confined light quarks, generating the nucleon scalar polarizability that drives nuclear saturation and enables a unified description of finite nuclei, dense nuclear matter, and neutron-star properties with only a few parameters, the same in-medium modification of quark structure also accounts naturally for the nuclear EMC  \cite{Guichon:2018uew}. Self-consistent nuclear densities are obtained within the relativistic Hartree formulation of the QMC model, where each nucleon occupies a Dirac orbital $\alpha$ generated within the model. The radial upper and lower components,
$G_\alpha(r)$ and $F_\alpha(r)$, of the corresponding spinor satisfy \cite{Guichon:1995ue, Mondal:2024vyt}
\begin{eqnarray}
\label{eq:KG}
\left(\frac{d}{dr}+\frac{\kappa}{r}\right)G_\alpha(r)
&=&
\bigl[m_N^*(r)-V(r)+E_\alpha\bigr]F_\alpha(r),
\nonumber\\
\left(\frac{d}{dr}-\frac{\kappa}{r}\right)F_\alpha(r)
&=&
\bigl[m_N^*(r)+V(r)-E_\alpha\bigr]G_\alpha(r),
\end{eqnarray}
where $E_\alpha$ denotes the full relativistic single-particle
eigen-energy and $\kappa=(-1)^{j+\ell+1/2}\left(j+\tfrac{1}{2}\right)$ is the Dirac quantum number encoding the angular momenta $(j,\ell)$ of the orbital.
The position-dependent Dirac effective mass is $m_N^*(r)=m_N-V_\sigma(r)-\frac{\tau^3}{2}V_\delta(r)$, where, in contrast to point-nucleon relativistic mean-field theory, the scalar potential $V_\sigma(r)=g_\sigma(\sigma)\,\sigma(r)$ carries a $\sigma$ field dependent coupling reflecting the self-consistent response of the confined quarks to the medium, the scalar polarizability of the nucleon, which is the defining feature of the QMC model.
The vector potential is
$V(r) = V_\omega(r) +\frac{\tau^3}{2}V_\rho(r) +\frac{1+\tau^3}{2}V_c(r)$, with $\tau^3=+1$ ($-1$) for protons (neutrons). The mesonic mean fields $M=\{\sigma,\delta,\omega,\rho\}$ satisfy the
inhomogeneous Klein--Gordon equations
\begin{equation}
\label{eq:KG-meson}
\left(
-\frac{d^2}{dr^2}
-\frac{2}{r}\frac{d}{dr}
+m_M^2
\right)M(r)
=
\mathcal{J}_M(r),
\end{equation}
with source terms $\mathcal{J}_M(r)$ constructed
self-consistently with the scalar and baryon densities within the model from the occupied nucleon orbitals. The Coulomb field $A(r)$ obeys the corresponding Poisson equation, obtained from Eq.~(\ref{eq:KG-meson}) in the massless limit $m_M\rightarrow 0$ with the proton charge density as source.
Equations~(\ref{eq:KG}) and (\ref{eq:KG-meson}) are iterated to self-consistency. The nucleon density is then built from the occupied orbitals,
\begin{equation}
\label{eq:den}
\rho(r)
=
\sum_{\alpha}^{\mathrm{occ}}
\frac{2j_\alpha+1}{4\pi r^2}
\left[
|G_\alpha(r)|^2+|F_\alpha(r)|^2
\right],
\end{equation}
with the radial wave functions normalized as in \cite{Mondal:2024vyt}
\begin{eqnarray}
    \int_0^\infty dr\,\big[|G_\alpha(r)|^2+|F_\alpha(r)|^2\big]=1
\end{eqnarray} with $4\pi\int_0^\infty dr\,r^2\rho(r)=A$.
The factor $2j_\alpha+1$ accounts for the magnetic degeneracy of a filled spherical shell. Details of the QMC interactions and parameters employed in this study are provided in the Supplemental Material \cite{SM}. Finally, the cross section is corrected for the real part of the scattering amplitude and for skewedness effects, as detailed in the Supplemental Material \cite{SM}.
\\

\paragraph*{Nuclear shell effects on vector-meson production.} 

\begin{figure}
	\centering
	\includegraphics[width=0.9\linewidth]{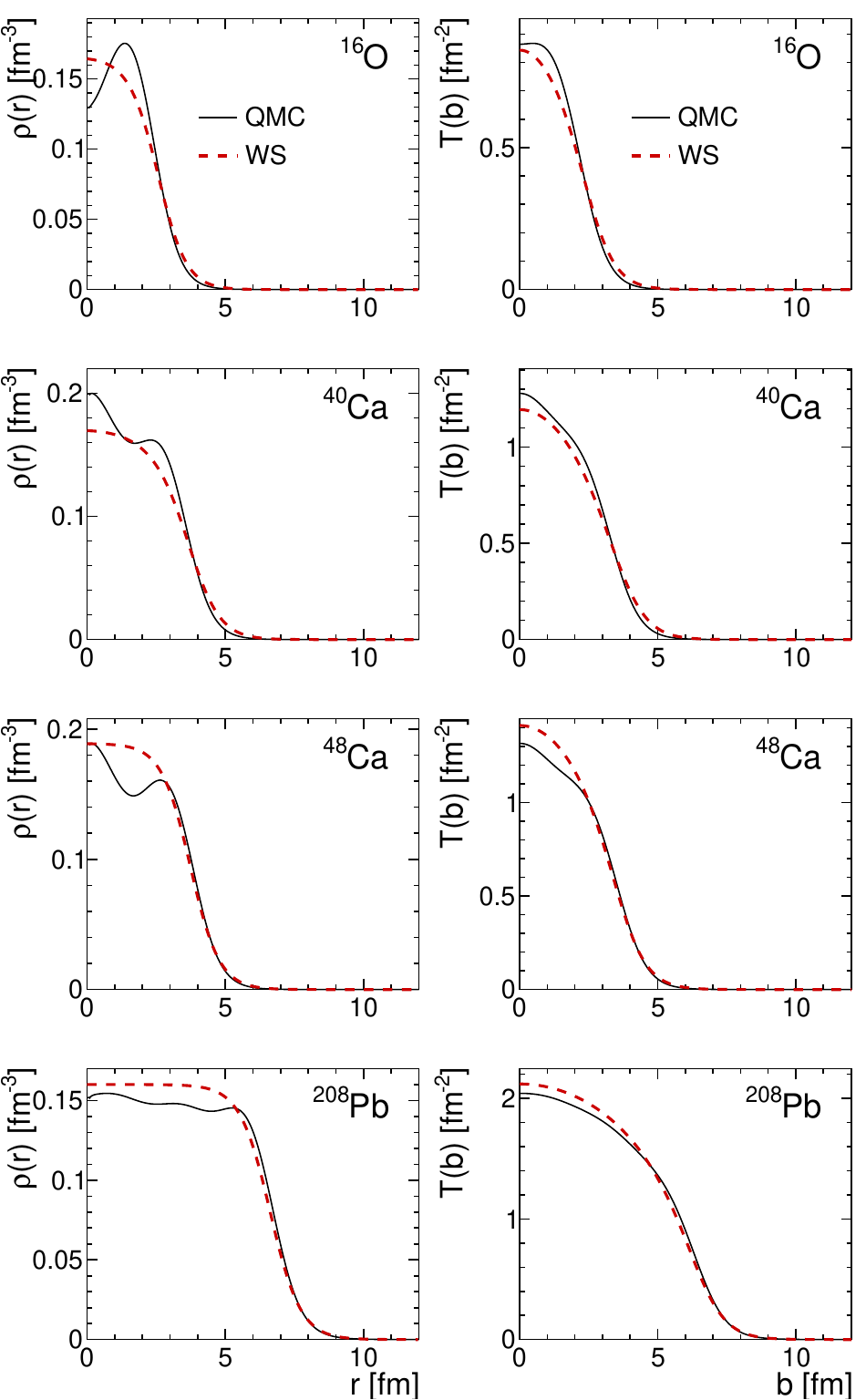}
	\caption{Nucleon density distributions $\rho(r)$ (left) and the corresponding thickness functions $T(b)$(right), for $\rm{^{16}O}$, $\rm{^{40}Ca}$, $\rm{^{48}Ca}$, and $\rm{^{208}Pb}$.
    The QMC-driven results (black, solid) are compared with the baseline Woods--Saxon parametrization (red, dashed).}
	\label{rho_T}
\end{figure}

~~~Fig.~\ref{rho_T} depicts the ground-state baryon density distributions $\rho(r) = \rho_p(r)+\rho_n(r)$, obtained self-consistently within the QMC framework, for \(^{16}\mathrm{O}\),
\(^{40}\mathrm{Ca}\), \(^{48}\mathrm{Ca}\), and
\(^{208}\mathrm{Pb}\) (left column), together with the corresponding thickness functions $T(b)$ (right column). In each panel the results are compared with the standard two-parameter Woods--Saxon (WS) distribution \cite{Woods:1954zz},
\begin{align}
\rho_{\rm WS}(r) = \frac{\rho_0}{1 + \exp\left[(r - R_0)/a\right]},
\end{align}
with radius $R_0$ and diffuseness $a$ taken from electron-scattering compilations.
\begin{figure}
	\centering
	\includegraphics[width=0.85\linewidth]{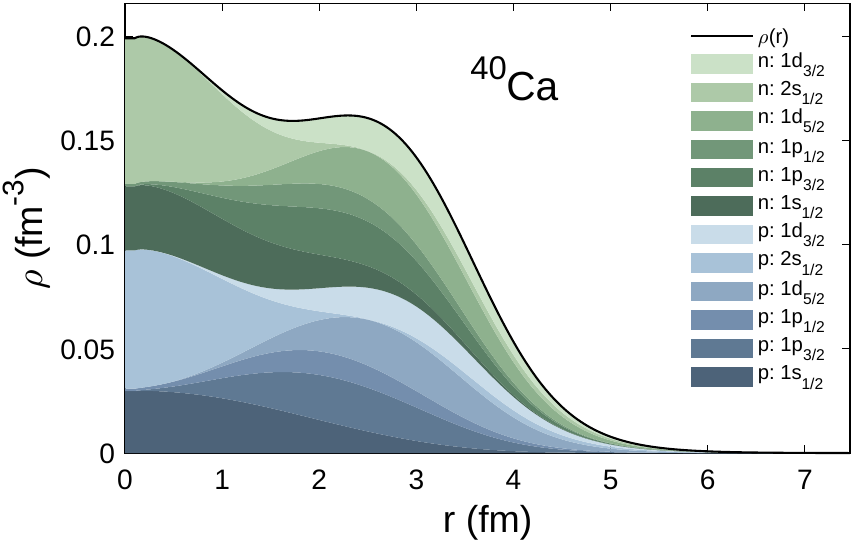}
	\caption{Orbital contributions to $^{40}\rm{Ca}$'s density profile.}
	\label{rho}
\end{figure}

\begin{figure}
	\centering
	\includegraphics[width=1.01\linewidth]{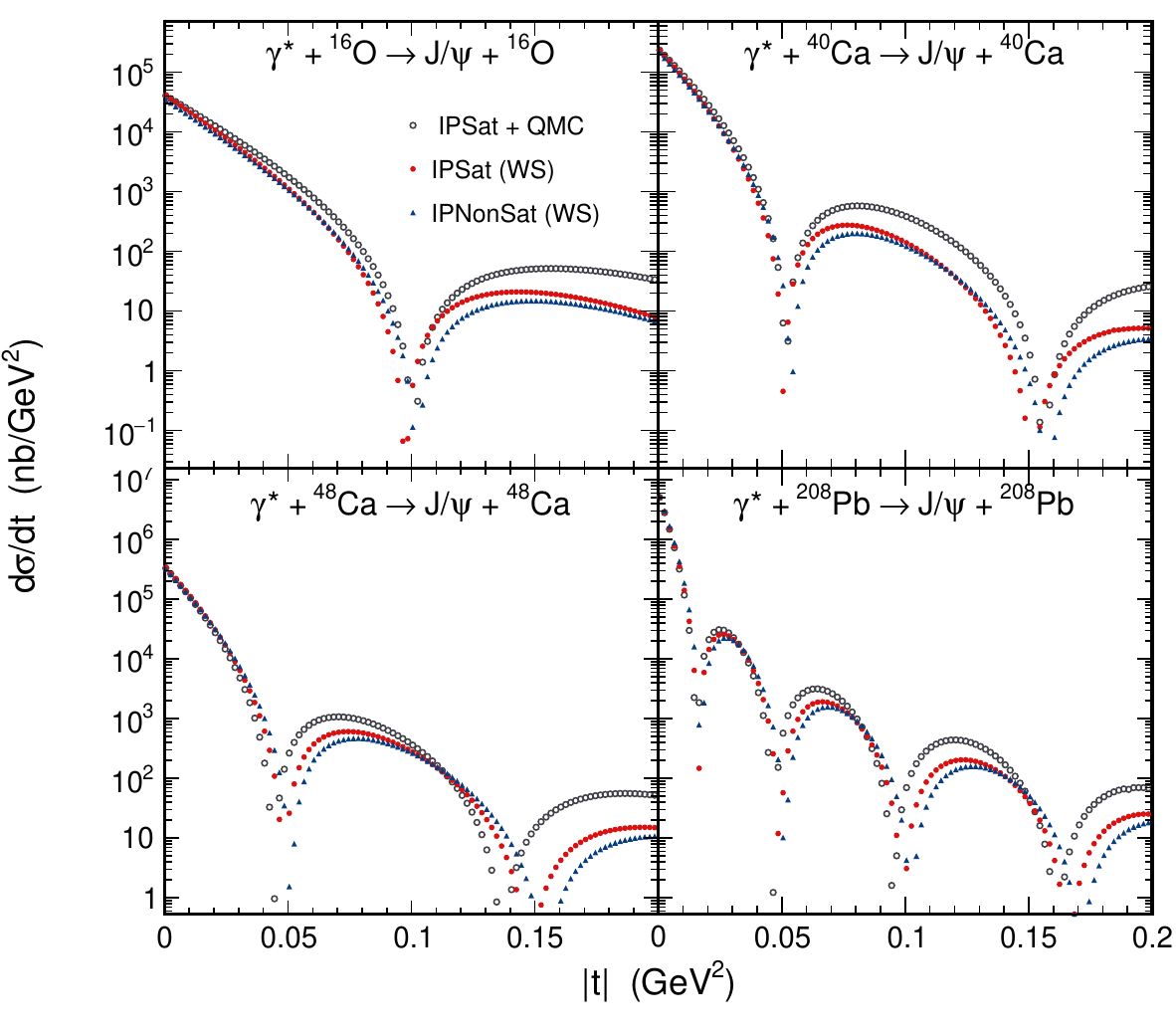}
	\caption{Coherent $J/\psi$ photo-production differential $t$-spectrum ($d\sigma^{\gamma^*A \rightarrow J/\psi\,A}/dt$). Results obtained with the QMC density profiles (IPSat + QMC) are compared with the Woods--Saxon baselines (WS) in IPSat and IPNonSat models.}
	\label{dsigmadt_Jpsi}
\end{figure}
By construction, the WS form is structureless and interpolates monotonically between a constant central density and an exponential surface falloff,
while the densities obtained in QMC are built from the squared Dirac wave functions of the occupied single-particle orbitals.
The core density follows directly from its single-particle contents, where $s$ orbitals are the only orbitals with non-vanishing amplitude at the origin and therefore fix the central density.
States with $l>0$ are suppressed at small $r$ and peak at intermediate radii, so the spin--orbit partner $d$ states dominate the surface region and govern the diffuseness.
Fig.~\ref{rho} illustrates this orbital-driven structure for $\rm{^{40}Ca}$. The Coulomb repulsion acting on protons shifts the proton orbitals slightly outward relative to their neutron partners, lowering the proton central density and enhancing its surface tail.
The deviation from the WS profile is largest near the center, and it decreases toward the surface, where both descriptions are constrained by the same charge radii \cite{Sick:1979bkt}. Light nuclei have a larger surface-to-volume ratio and fewer occupied shells, so the modulations are stronger in \(\rho(r)\).

The projection onto the transverse plane partially averages the shell-induced modulations of the three-dimensional density. The resulting thickness functions, shown in the right column of
Fig.~(\ref{rho_T}), nevertheless retain a visible imprint, whose relative magnitude decreases with increasing nuclear mass. This follows from a separation of scales, shell oscillations vary on the Fermi-wavelength scale, whereas the longitudinal path length grows as \(\sim A^{1/3}\). A localized density modulation therefore contributes a progressively smaller fraction of the integrated thickness
in heavy nuclei, while in light nuclei the same spatial structure occupies a larger fraction of \(T(b)\) and yields a stronger relative modulation. 

Fig.~\ref{dsigmadt_Jpsi} shows the coherent $|t|$-differential cross sections for \(J/\psi\) photoproduction at \(W=75\,\mathrm{GeV}\) for \(^{16}\mathrm{O}\), \(^{40}\mathrm{Ca}\), \(^{48}\mathrm{Ca}\), and \(^{208}\mathrm{Pb}\) in the saturation model with shell-structure densities (IPSat+ QMC), and in IPSat and IPNonSat with a Woods--Saxon baseline.
Because of the small dipole size of the \(J/\psi\), saturation effects are negligible and the IPSat and IPNonSat Woods--Saxon curves nearly
coincide, rendering \(J/\psi\) production particularly well suited
for nuclear spectroscopy.

For oxygen, the QMC profile leaves the position of the first minimum ($|t|\simeq 0.10GeV^2$) essentially unchanged but enhances the cross section throughout the first cone and, more significantly, raises the height of the secondary maximum relative to the Woods--Saxon baseline.

To isolate the effects associated with the microscopic nuclear density, we examine the $t$-spectrum for two calcium isotopes, $^{40}\mathrm{Ca}$ and $^{48}\mathrm{Ca}$. Recently, It was also measured at JLab experiments  \cite{Nguyen2026NuclearShellStructure, Nguyen:2026ztp} that the short range correlations are governed by the long range shell structure, so we expect that any density effects would be visible in these spectra.  The modifications of the coherent spectrum exhibit a clear isotopic dependence, as expected from the different shell occupancies of the two nuclei. In particular, the higher-order diffractive minima are displaced in opposite directions. For $^{40}\mathrm{Ca}$, the second minimum shifts toward larger $|t|$, whereas for $^{48}\mathrm{Ca}$ it shifts toward smaller $|t|$. Since the proton number is unchanged, the comparison between $^{40}\mathrm{Ca}$ and $^{48}\mathrm{Ca}$ isolates the effect of the eight additional neutrons occupying the $f_{7/2}$ shell in $^{48}\mathrm{Ca}$. The opposite shifts of the diffractive minima therefore demonstrate a direct sensitivity of the coherent spectrum to the underlying shell structure and the corresponding microscopic density distribution. For both cases, the heights of the second and higher-order maxima are enhanced, although the magnitude of the enhancement differs between the two isotopes.

For $^{208}\mathrm{Pb}$, the first diffractive minima of all three curves essentially coincide. The QMC density predominantly enhances the magnitudes of the diffractive lobes, while the residual differences increase gradually with the order of the minimum and become appreciable only beyond the second diffractive minimum.

In experiment, the $t$-spectrum encodes a broad class of nuclear
effects. Light nuclei remain most sensitive to these contributions, while
many-body correlations such as clustering are expected to weaken with increasing mass and have been shown to be already negligible for neon
($A=20$)~\cite{abintio}.
Our predictions therefore provide a baseline for isolating the contribution of occupied single-particle orbitals and for estimating residual many-body correlations.
Calcium emerges as a promising intermediate-mass candidate for probing nuclear shell structure.

\begin{figure}
	\centering
	\includegraphics[width=1.01\linewidth]{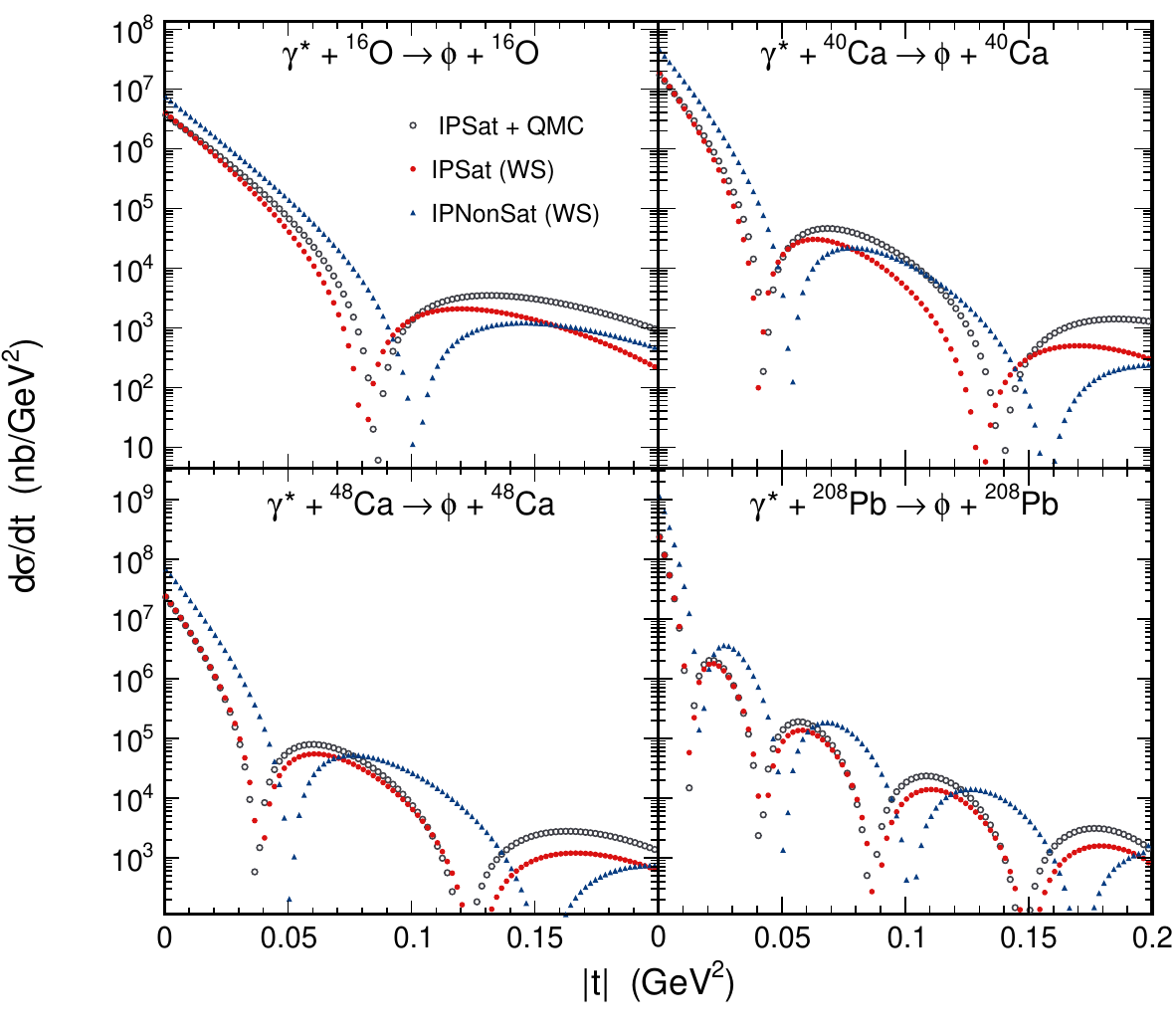}
	\caption{Coherent $\phi$ photo-production differential $t$-spectrum ($d\sigma^{\gamma^*A \rightarrow \phi\,A}/dt$). Results obtained with the QMC density profiles (IPSat + QMC) are compared with the Woods--Saxon baselines (WS) in IPSat and IPNonSat models.}
	\label{dsigmadt_phi}
\end{figure}

Figure~\ref{dsigmadt_phi} shows the coherent $|t|$-differential cross sections for \(\phi\) photoproduction at \(W=75\,\mathrm{GeV}\) off \(^{16}\mathrm{O}\),
\(^{40}\mathrm{Ca}\), \(^{48}\mathrm{Ca}\), and \(^{208}\mathrm{Pb}\) in the IPSat model with shell-structure densities, and in IPSat and IPNonSat with a Woods--Saxon baseline.
Owing to the larger characteristic dipole size probed in $\phi$ production, the $|t|$-spectrum exhibits enhanced sensitivity to nonlinear QCD dynamics. Consequently, the linearized IPNonSat prediction separates significantly from the saturated IPSat result, producing a larger cross section at low $|t|$ and in the higher-order diffractive lobes, while also shifting the positions of the diffractive minima across the full range of nuclear masses considered. In contrast to the $J/\psi$ channel, shell-structure effects constitute a sub-leading modification relative to non-linear effects in $\phi$ production. Nevertheless, their qualitative impact on the $t$-spectrum remains similar to that observed for $J/\psi$, providing a controlled baseline against which saturation-induced modifications can be quantified. Consequently, coherent $\phi$ production offers a comparatively clean probe of nonlinear saturation dynamics, particularly for heavy nuclei, where shell-structure effects on the diffractive spectrum are relatively small.
\begin{figure}
	\centering
	\includegraphics[width=0.85\linewidth]{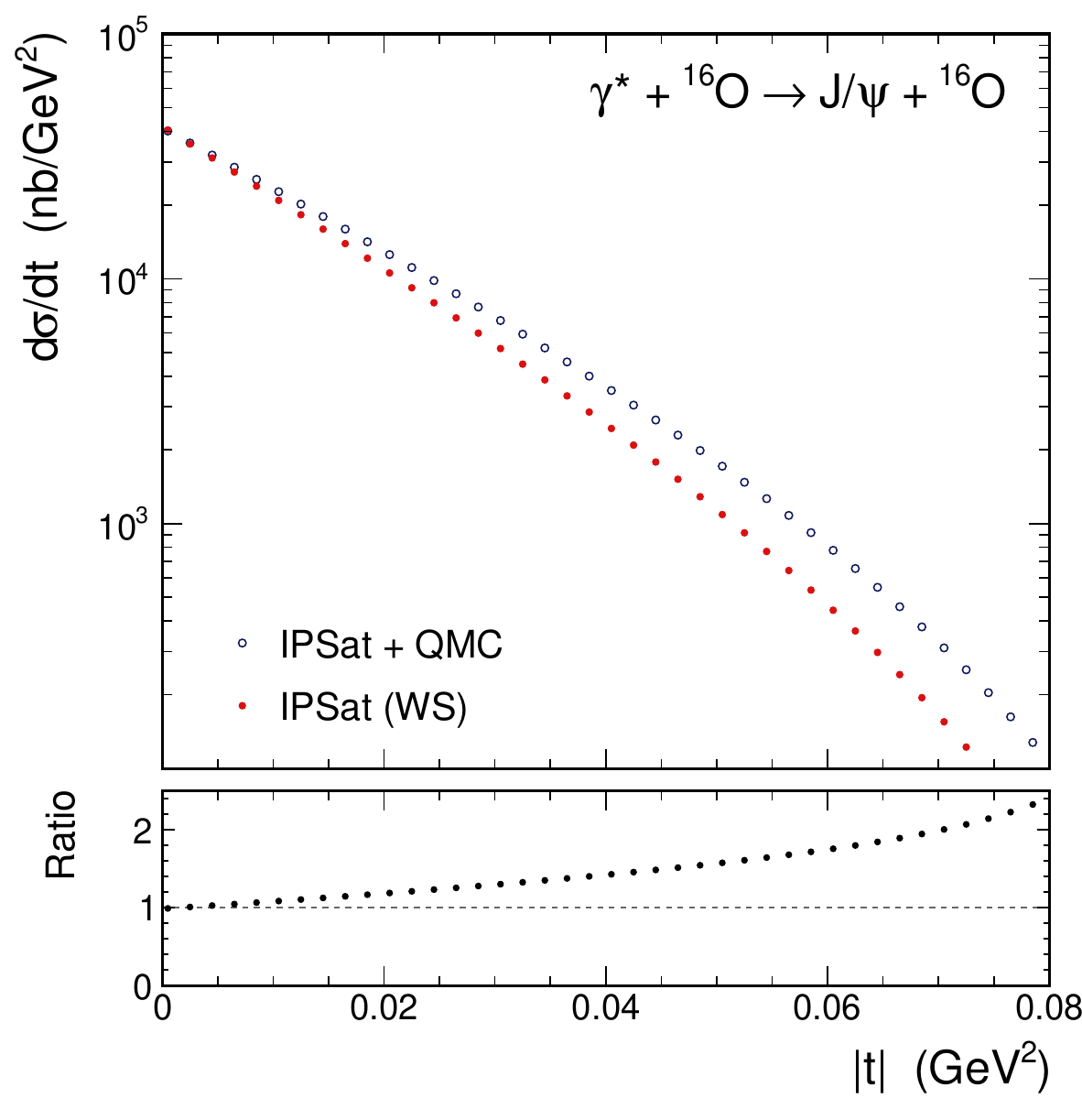}
	\caption{Differential cross section $\mathrm{d}\sigma/\mathrm{d}t$ for exclusive $J/\psi$ photoproduction off $^{16}\text{O}$ as a function of $\vert{}t\vert{}$ (top panel), compared with the Woods-Saxon (WS) nuclear profile. The lower panel displays the ratio of the two predictions.}
	\label{dsigmadt_phi_2}
\end{figure}
Figure~\ref{dsigmadt_phi_2} shows a magnified view of $J/\psi$ photo-production off oxygen, together with the ratio of the IPSat+QMC cross section to the IPSat Woods--Saxon baseline. 
The ratio peaks near the diffractive minima, where a small displacement of the dip between the two geometries produces a local enhancement, reaching the ratio of the order $\sim 2$, corresponding to a two-fold enhancement in cross section, underscoring the impact of nuclear shell structure on the coherent $t$ spectrum.
If such spectra can be measured with sufficient precision at the EIC, the diffraction pattern in nuclear isotopes for intermediate-mass nuclei can be used to probe nuclear shell structure where the differences in spectrum would be dictated by the shell densities .\\

\paragraph*{Summary and Outlook.}
We have shown that coherent exclusive vector-meson production in $eA$ collisions retains an imprint of microscopic nuclear shell structure.
The nuclear densities are obtained self-consistently in the QMC model, so that radial variations map onto occupied shells within a unified framework that describes finite nuclei and dense matter, including neutron stars.
Our results have three main consequences.
First, the $|t|$-differential \(J/\psi\) photoproduction cross section on intermediate-mass nuclei is a sensitive probe of nuclear shell structure, and the nuclear isotopes such as $^{40}\text{Ca}$--$^{48}\text{Ca}$ pair further isolates the effect of shell filling at fixed proton number.
Second, for light nuclei the same observable provides a controlled baseline from which residual many-body correlations can be isolated and quantified.
Third, $\phi$ photoproduction on light and intermediate nuclei requires shell structure in the initial-state geometry before nonlinear effects can be isolated, whereas heavy nuclei are largely insensitive to these modulations over the first few diffraction minima and can therefore directly probe saturation dynamics.

Our results quantify how much of the observable $t$-structure is already generated by mean-field-driven single-particle orbitals across the nuclear chart, thereby establishing a crucial link between a fully correlated many-body system and an independent-particle description in nuclear QCD.
A comparison of the saturation model employing QMC densities with measurements of coherent $J/\psi$ photo production in ultra peripheral Pb--Pb collisions is presented in the Supplemental Material \cite{SM}, showing good agreement with the experimental data \cite{ALICE:2023gcs}.

The knowledge of the nuclear initial state is complementary to that
inferred from heavy-ion collisions and is an essential input to
hydrodynamic modeling. A natural extension of our work is to isolate the contribution of nuclear shell structure to the incoherent cross section for both spherical and deformed nuclei.

\paragraph*{Acknowledgments.}
We thank Jaswant Singh for useful discussions.
The work of A.M is supported by the IPDF grant at the Indian Institute of Technology Bombay (Seed Grant No. RD/0525-IRCCSH0-033).
A.K is supported by Center for Frontiers in Nuclear Science.

\bibliographystyle{elsarticle-num}
\bibliography{bibliography}

@article{Nguyen2026NuclearShellStructure,
  author  = {Nguyen, D. and others},
  title   = {Nuclear shell structure governs short-range nucleon pairing},
  journal = {Nature},
  year    = {2026},
  volume  = {654},
  number  = {8119},
  pages   = {619--621},
  doi     = {10.1038/s41586-026-10616-2}
}

@article{Nguyen:2026ztp,
    author = "Nguyen, D. and others",
    title = "{Short-range correlated pair formation and nuclear shell structure}",
    eprint = "2606.07754",
    archivePrefix = "arXiv",
    primaryClass = "nucl-ex",
    month = "6",
    year = "2026"
}

@misc{SM,
    note = { Supplemental Material:Imprints of nuclear shell structure in exclusive vector meson production at [URL will be inserted by publisher]}
}

@article{Kumar:2022aly,
    author = "Kumar, Arjun and Toll, Tobias",
    title = "{Energy dependence of the proton geometry in exclusive vector meson production}",
    eprint = "2202.06631",
    archivePrefix = "arXiv",
    primaryClass = "hep-ph",
    doi = "10.1103/PhysRevD.105.114011",
    journal = "Phys. Rev. D",
    volume = "105",
    number = "11",
    pages = "114011",
    year = "2022"
}

@article{QMC1,
  title = {Quark Structure and Nuclear Effective Forces},
  author = {Guichon, P. A. M. and Thomas, A. W.},
  journal = {Phys. Rev. Lett.},
  volume = {93},
  issue = {13},
  pages = {132502},
  numpages = {4},
  year = {2004},
  month = {Sep},
  publisher = {American Physical Society},
  doi = {10.1103/PhysRevLett.93.132502}
}

@article{QMC2,
  title = {Finite Nuclei in the Quark-Meson Coupling Model},
  author = {Stone, J. R. and Guichon, P. A. M. and Reinhard, P. G. and Thomas, A. W.},
  journal = {Phys. Rev. Lett.},
  volume = {116},
  issue = {9},
  pages = {092501},
  numpages = {5},
  year = {2016},
  month = {Feb},
  publisher = {American Physical Society},
  doi = {10.1103/PhysRevLett.116.092501}
}

@article{FSgeom1,
  title = {Initial-state geometry and fluctuations in Au + Au, Cu + Au, and U + U collisions at energies available at the BNL Relativistic Heavy Ion Collider},
  author = {Schenke, Bj\"orn and Tribedy, Prithwish and Venugopalan, Raju},
  journal = {Phys. Rev. C},
  volume = {89},
  issue = {6},
  pages = {064908},
  numpages = {10},
  year = {2014},
  month = {Jun},
  publisher = {American Physical Society},
  doi = {10.1103/PhysRevC.89.064908}
}

@article{FSgeom2,
  title = {Impact of Nuclear Deformation on Relativistic Heavy-Ion Collisions: Assessing Consistency in Nuclear Physics across Energy Scales},
  author = {Giacalone, Giuliano and Jia, Jiangyong and Zhang, Chunjian},
  journal = {Phys. Rev. Lett.},
  volume = {127},
  issue = {24},
  pages = {242301},
  numpages = {6},
  year = {2021},
  month = {Dec},
  publisher = {American Physical Society},
  doi = {10.1103/PhysRevLett.127.242301}
}

@article{FSgeom3,
  title = {Evidence of Quadrupole and Octupole Deformations in $^{96}\mathrm{Zr}+^{96}\mathrm{Zr}$ and $^{96}\mathrm{Ru}+^{96}\mathrm{Ru}$ Collisions at Ultrarelativistic Energies},
  author = {Zhang, Chunjian and Jia, Jiangyong},
  journal = {Phys. Rev. Lett.},
  volume = {128},
  issue = {2},
  pages = {022301},
  numpages = {6},
  year = {2022},
  month = {Jan},
  publisher = {American Physical Society},
  doi = {10.1103/PhysRevLett.128.022301}
}

@article{FSgeom4,
  title = {Evidence of Hexadecapole Deformation in Uranium-238 at the Relativistic Heavy Ion Collider},
  author = {Ryssens, Wouter and Giacalone, Giuliano and Schenke, Bj\"orn and Shen, Chun},
  journal = {Phys. Rev. Lett.},
  volume = {130},
  issue = {21},
  pages = {212302},
  numpages = {7},
  year = {2023},
  month = {May},
  publisher = {American Physical Society},
  doi = {10.1103/PhysRevLett.130.212302}
}

@article{lightnuclei,
  title = {Exclusive production of light vector mesons at next-to-leading order in the dipole picture},
  author = {M\"antysaari, Heikki and Penttala, Jani},
  journal = {Phys. Rev. D},
  volume = {105},
  issue = {11},
  pages = {114038},
  numpages = {22},
  year = {2022},
  month = {Jun},
  publisher = {American Physical Society},
}

@article{multiscale,
  title = {Multiscale Imaging of Nuclear Deformation at the Electron-Ion Collider},
  author = {M\"antysaari, Heikki and Schenke, Bj\"orn and Shen, Chun and Zhao, Wenbin},
  journal = {Phys. Rev. Lett.},
  volume = {131},
  issue = {6},
  pages = {062301},
  numpages = {7},
  year = {2023},
  month = {Aug},
  publisher = {American Physical Society},

 
}

@article{abintio,
  title = {Nuclear structure and saturation effects from diffractive vector meson production},
  author = {M\"antysaari, Heikki and Roch, Hendrik and Schenke, Bj\"orn and Shen, Chun and Zhao, Wenbin},
  journal = {Phys. Rev. D},
  volume = {114},
  issue = {1},
  pages = {014068},
  numpages = {15},
  year = {2026},
  month = {Jul},
  publisher = {American Physical Society},
 

}

@article{AbdulKhalek:2021gbh,
	author = "Abdul Khalek, R. and others",
	title = "{Science Requirements and Detector Concepts for the Electron-Ion Collider: EIC Yellow Report}",
	eprint = "2103.05419",
	archivePrefix = "arXiv",
	primaryClass = "physics.ins-det",
	reportNumber = "BNL-220990-2021-FORE, JLAB-PHY-21-3198, LA-UR-21-20953",
	month = "3",
	year = "2021"
}

@article{Accardi:2012qut,
	author = "Accardi, A. and others",
	editor = "Deshpande, A. and Meziani, Z. E. and Qiu, J. W.",
	title = "{Electron Ion Collider: The Next QCD Frontier}: {Understanding the glue that binds us all}",
	eprint = "1212.1701",
	archivePrefix = "arXiv",
	primaryClass = "nucl-ex",
	reportNumber = "BNL-98815-2012-JA, JLAB-PHY-12-1652",
	doi = "10.1140/epja/i2016-16268-9",
	journal = "Eur. Phys. J. A",
	volume = "52",
	number = "9",
	pages = "268",
	year = "2016"
}

@article{Kumar:2021zbn,
	author = "Kumar, Arjun and Toll, Tobias",
	title = "{Investigating the structure of gluon fluctuations in the proton with incoherent diffraction at HERA}",
	eprint = "2106.12855",
	archivePrefix = "arXiv",
	primaryClass = "hep-ph",
	month = "6",
	year = "2021"
}

@article{Woods:1954zz,
    author = "Woods, Roger D. and Saxon, David S.",
    title = "{Diffuse Surface Optical Model for Nucleon-Nuclei Scattering}",
    doi = "10.1103/PhysRev.95.577",
    journal = "Phys. Rev.",
    volume = "95",
    pages = "577--578",
    year = "1954"
}

@article{Sick:1979bkt,
    author = "Sick, I. and Bellicard, J. B. and Cavedon, J. M. and Frois, B. and Huet, M. and Leconte, P. and Phan, X. H. and Platchkov, S.",
    title = "{CHARGE DENSITY OF CA-40}",
    doi = "10.1016/0370-2693(79)90458-1",
    journal = "Phys. Lett. B",
    volume = "88",
    pages = "245--248",
    year = "1979"
}

@article{Guichon:1995ue,
    author = "Guichon, Pierre A. M. and Saito, Koichi and Rodionov, Evgenii N. and Thomas, Anthony William",
    title = "{The Role of nucleon structure in finite nuclei}",
    eprint = "nucl-th/9509034",
    archivePrefix = "arXiv",
    reportNumber = "ADP-95-45-T-194, DAPNIA-SPHN-95-51",
    doi = "10.1016/0375-9474(96)00033-4",
    journal = "Nucl. Phys. A",
    volume = "601",
    pages = "349--379",
    year = "1996"
}

@article{Mondal:2024vyt,
    author = "Mondal, Arpita and Mishra, Amruta",
    title = "{Meson-nucleus bound states in the quark meson coupling model}",
    eprint = "2407.19896",
    archivePrefix = "arXiv",
    primaryClass = "nucl-th",
    doi = "10.1103/PhysRevC.110.055201",
    journal = "Phys. Rev. C",
    volume = "110",
    number = "5",
    pages = "055201",
    year = "2024",
    note = "[Erratum: Phys.Rev.C 112, 019901 (2025)]"
}

@article{Santos:2009yj,
    author = "Santos, Alexandre M. and Panda, Prafulla K. and Providencia, Constanca",
    title = "{Low density instabilities in asymmetric nuclear matter within QMC with delta-meson}",
    eprint = "0901.3243",
    archivePrefix = "arXiv",
    primaryClass = "nucl-th",
    doi = "10.1103/PhysRevC.79.045805",
    journal = "Phys. Rev. C",
    volume = "79",
    pages = "045805",
    year = "2009"
}

@article{Kowalski:2003hm,
author         = "Kowalski, Henri and Teaney, Derek",
title          = "{An Impact parameter dipole saturation model}",
journal        = "Phys. Rev.",
volume         = "D68",
year           = "2003",
pages          = "114005",
doi            = "10.1103/PhysRevD.68.114005",
eprint         = "hep-ph/0304189",
archivePrefix  = "arXiv",
primaryClass   = "hep-ph",
SLACcitation   = "%%CITATION = HEP-PH/0304189;%%"
}

@article{Schenke:2012hg,
author         = "Schenke, Björn and Tribedy, Prithwish and Venugopalan,
Raju",
title          = "{Event-by-event gluon multiplicity, energy density, and
	eccentricities in ultrarelativistic heavy-ion collisions}",
journal        = "Phys. Rev.",
volume         = "C86",
pages          = "034908",
doi            = "10.1103/PhysRevC.86.034908",
year           = "2012",
eprint         = "1206.6805",
archivePrefix  = "arXiv",
primaryClass   = "hep-ph",
SLACcitation   = "%%CITATION = ARXIV:1206.6805;%%",
}

@article{Schenke:2012wb,
author         = "Schenke, Björn and Tribedy, Prithwish and Venugopalan,
Raju",
title          = "{Fluctuating Glasma initial conditions and flow in heavy
	ion collisions}",
journal        = "Phys. Rev. Lett.",
volume         = "108",
pages          = "252301",
doi            = "10.1103/PhysRevLett.108.252301",
year           = "2012",
eprint         = "1202.6646",
archivePrefix  = "arXiv",
primaryClass   = "nucl-th",
SLACcitation   = "%%CITATION = ARXIV:1202.6646;%%",
}

@article{Nikolaev:1990ja,
author         = "Nikolaev, Nikolai N. and Zakharov, B. G.",
title          = "{Color transparency and scaling properties of nuclear
	shadowing in deep inelastic scattering}",
journal        = "Z. Phys.",
volume         = "C49",
year           = "1991",
pages          = "607-618",
doi            = "10.1007/BF01483577",
reportNumber   = "OUTP-90-23-P",
SLACcitation   = "%%CITATION = ZEPYA,C49,607;%%"
}

@article{Kowalski:2006hc,
author         = "Kowalski, H. and Motyka, L. and Watt, G.",
title          = "{Exclusive diffractive processes at HERA within the
	dipole picture}",
journal        = "Phys. Rev.",
volume         = "D74",
year           = "2006",
pages          = "074016",
doi            = "10.1103/PhysRevD.74.074016",
eprint         = "hep-ph/0606272",
SLACcitation   = "%%CITATION = HEP-PH/0606272;%%"
}

@article{Mueller:1993rr,
author         = "Mueller, Alfred H.",
title          = "{Soft gluons in the infinite momentum wave function and
	the BFKL pomeron}",
journal        = "Nucl. Phys.",
volume         = "B415",
year           = "1994",
pages          = "373-385",
doi            = "10.1016/0550-3213(94)90116-3",
reportNumber   = "SLAC-PUB-10047, CU-TP-609",
SLACcitation   = "%%CITATION = NUPHA,B415,373;%%"
}

@article{Chekanov:2002xi,
author         = "Chekanov, S. and others",
title          = "{Exclusive photoproduction of J / psi mesons at HERA}",
collaboration  = "ZEUS",
journal        = "Eur. Phys. J.",
volume         = "C24",
year           = "2002",
pages          = "345-360",
doi            = "10.1007/s10052-002-0953-7",
eprint         = "hep-ex/0201043",
archivePrefix  = "arXiv",
primaryClass   = "hep-ex",
reportNumber   = "DESY-02-008",
SLACcitation   = "%%CITATION = HEP-EX/0201043;%%"
}

@article{Aktas:2005xu,
author         = "Aktas, A. and others",
title          = "{Elastic J/psi production at HERA}",
collaboration  = "H1",
journal        = "Eur. Phys. J.",
volume         = "C46",
year           = "2006",
pages          = "585-603",
doi            = "10.1140/epjc/s2006-02519-5",
eprint         = "hep-ex/0510016",
archivePrefix  = "arXiv",
primaryClass   = "hep-ex",
reportNumber   = "DESY-05-161",
SLACcitation   = "%%CITATION = HEP-EX/0510016;%%"
}

@article{Shuvaev:1999ce,
	title        = {{Off diagonal distributions fixed by diagonal partons at small x and xi}},
	author       = {Shuvaev, A. G. and Golec-Biernat, Krzysztof J. and Martin, Alan D. and Ryskin, M. G.},
	year         = 1999,
	journal      = {Phys. Rev. D},
	volume       = 60,
	pages        = {014015},
	doi          = {10.1103/PhysRevD.60.014015},
	eprint       = {hep-ph/9902410},
	archiveprefix = {arXiv},
	reportnumber = {DTP-99-18}
}

@article{Lappi:2010dd,
	author = "Lappi, T. and Mantysaari, H.",
	title = "{Incoherent diffractive J/Psi-production in high energy nuclear DIS}",
	eprint = "1011.1988",
	archivePrefix = "arXiv",
	primaryClass = "hep-ph",
	reportNumber = "INT-PUB-10-061",
	doi = "10.1103/PhysRevC.83.065202",
	journal = "Phys. Rev. C",
	volume = "83",
	pages = "065202",
	year = "2011"
}

@article{Guichon:1987jp,
    author = "Guichon, Pierre A. M.",
    title = "{A Possible Quark Mechanism for the Saturation of Nuclear Matter}",
    reportNumber = "LYCEN-8762",
    doi = "10.1016/0370-2693(88)90762-9",
    journal = "Phys. Lett. B",
    volume = "200",
    pages = "235--240",
    year = "1988"
}

@article{Saito:2005rv,
    author = "Saito, K. and Tsushima, Kazuo and Thomas, Anthony William",
    title = "{Nucleon and hadron structure changes in the nuclear medium and impact on observables}",
    eprint = "hep-ph/0506314",
    archivePrefix = "arXiv",
    reportNumber = "JLAB-THY-05-326",
    doi = "10.1016/j.ppnp.2005.07.003",
    journal = "Prog. Part. Nucl. Phys.",
    volume = "58",
    pages = "1--167",
    year = "2007"
}

@article{Guichon:2018uew,
    author = "Guichon, P. A. M. and Stone, J. R. and Thomas, A. W.",
    title = "{Quark{\textendash}Meson-Coupling (QMC) model for finite nuclei, nuclear matter and beyond}",
    eprint = "1802.08368",
    archivePrefix = "arXiv",
    primaryClass = "nucl-th",
    reportNumber = "ADP-18-5/T1053, IRFU-18-03, ADP-18-5-T1053",
    doi = "10.1016/j.ppnp.2018.01.008",
    journal = "Prog. Part. Nucl. Phys.",
    volume = "100",
    pages = "262--297",
    year = "2018"
}

@article{ZEUS:2007iet,
	author = "Chekanov, S. and others",
	collaboration = "ZEUS",
	title = "{Exclusive rho0 production in deep inelastic scattering at HERA}",
	eprint = "0708.1478",
	archivePrefix = "arXiv",
	primaryClass = "hep-ex",
	reportNumber = "DESY-07-118",
	doi = "10.1186/1754-0410-1-6",
	journal = "PMC Phys. A",
	volume = "1",
	pages = "6",
	year = "2007"
}

@article{Toll:2012mb,
	title        = {{Exclusive diffractive processes in electron-ion collisions}},
	author       = {Toll, Tobias and Ullrich, Thomas},
	year         = 2013,
	journal      = {Phys. Rev. C},
	volume       = 87,
	number       = 2,
	pages        = {024913},
	doi          = {10.1103/PhysRevC.87.024913},
	eprint       = {1211.3048},
	archiveprefix = {arXiv},
	primaryclass = {hep-ph},
	slaccitation = {%%CITATION = ARXIV:1211.3048;%%}
}

@article{Toll:2013gda,
	title        = {{The dipole model Monte Carlo generator Sar$t$re 1}},
	author       = {Toll, Tobias and Ullrich, Thomas},
	year         = 2014,
	journal      = {Comput. Phys. Commun.},
	volume       = 185,
	pages        = {1835--1853},
	doi          = {10.1016/j.cpc.2014.03.010},
	eprint       = {1307.8059},
	archiveprefix = {arXiv},
	primaryclass = {hep-ph},
	slaccitation = {%%CITATION = ARXIV:1307.8059;%%}
}

@article{ALICE:2023gcs,
    author = "Acharya, Shreyasi and others",
    collaboration = "ALICE",
    title = "{First Measurement of the |t| Dependence of Incoherent J/{\ensuremath{\psi}} Photonuclear Production}",
    eprint = "2305.06169",
    archivePrefix = "arXiv",
    primaryClass = "nucl-ex",
    reportNumber = "CERN-EP-2023-080",
    doi = "10.1103/PhysRevLett.132.162302",
    journal = "Phys. Rev. Lett.",
    volume = "132",
    number = "16",
    pages = "162302",
    year = "2024"
}
\clearpage

\renewcommand\a{\alpha}
\renewcommand\b{\beta}
\renewcommand\d{\delta}
\renewcommand\i{\iota}
\renewcommand\k{\kappa}
\renewcommand\l{\lambda}
\renewcommand\t{\tau}
\renewcommand\u{\upsilon}
\renewcommand\c{\chi}
\renewcommand\j{\psi}
\renewcommand\o{\omega}

\renewcommand\L{\Lambda}


\clearpage
\onecolumngrid
\begin{center}
{\Large\bfseries
Supplemental Material: Imprints of Nuclear Shell Structure in Exclusive Vector Meson Production
\par}
\vspace{1.2em}
{\large
Arpita Mondal,$^{1,*}$
Arjun Kumar,$^{2,*}$
and Debojit Sarkar$^{1}$
\par}
\vspace{0.8em}
{\small
$^{1}$Department of Physics,
Indian Institute of Technology Bombay,
Mumbai-400076, India
\par}
\vspace{0.4em}
{\small
$^{2}$Center for Frontiers in Nuclear Science,
Department of Physics and Astronomy,
Stony Brook University,
New York 11794--3800, USA
\par}
\end{center}
\vspace{1em}
\twocolumngrid

\paragraph*{The QMC model.}
The quark--meson coupling (QMC) model describes the low-energy nuclear interaction at the quark level: the scalar ($\s$, $\d$) and vector ($\o$, $\rho$) meson fields couple directly to the confined quarks of the nucleons, and the nuclear binding emerges from the self-consistent response of the nucleon internal structure to these fields.
At the hadronic level, the dynamics is governed by the Lagrangian density \cite{Mondal:2024vyt}
\begin{align}
\label{A1}
    \mathcal{L}  &=  \bar{\psi}_N\Big[i\g.\partial
    - \Big(m_N - \tilde{g}_\s(\s)\,\s
    - \tilde{g}_\d(\d)\tfrac{\tau^a}{2}\d^a\Big)\nonumber\\
    & - \g^\mu\Big(g_\o \o_\mu + g_\rho\tfrac{\tau^a}{2}\rho_{\mu}^a
    + \tfrac{e}{2}(1+\tau^a)A_{\mu}\Big)\Big]\psi_N \nonumber\\
    &+\tfrac{1}{2}\big(\partial_\mu\s\,\partial^\mu\s - m_\s^2\s^2\big)
    + \tfrac{1}{2}\big(\partial_\mu\d^a\partial^\mu\d^a
    - m_\d^2\d^a\d^a\big)\nonumber\\
    & - \Big[\tfrac{1}{4}\o_{\mu\nu}\o^{\mu\nu}
    - \tfrac{1}{2}m_\o^2\o_\mu\o^\mu\Big]\nonumber\\
    &- \Big(\tfrac{1}{4}\rho_{\mu\nu}^a\rho^{\mu\nu,a}
    - \tfrac{1}{2}m_\rho^2\rho_{\mu}^a\rho^{\mu,a}\Big)
    - \tfrac{1}{4}A_{\mu\nu}A^{\mu\nu},
\end{align}

where $\bar{\psi}_N = (\psi_p ~~ \psi_n)$ is the nucleon isodoublet, $m_N$, $m_\s$, $m_\o$, $m_\d$, and $m_\rho$ are the nucleon and meson masses, $\tau^a/2$ ($a=1,2,3$) is the nucleon isospin operator, and $\o_{\mu\nu}$, $\rho_{\mu\nu}^a$, and $A_{\mu\nu}$ are the field-strength tensors of the vector meson and photon fields.
The distinguishing feature of the QMC framework relative to purely hadronic relativistic mean-field models is that the scalar couplings $\tilde{g}_\s(\s)$ and $\tilde{g}_\d(\d)$ are not independent functions but are \emph{derived} from the quark structure of the nucleon: the effective nucleon mass $m_N^\star(\s,\d)$ is identified with the energy of a static MIT bag whose quarks move in the local mean fields, and the field-dependent couplings follow from the scalar response of the bag.
The couplings $g_\s^q$, $g_\o^q$, $g_\rho^q$ and $g_\d^q$ are fitted to the nuclear bulk properties \cite{Santos:2009yj}.
The nucleon-level couplings entering Eq.~(\ref{A1}) follow from the quark-level ones. The resulting charge radii of the four nuclei and the nuclear-matter bulk properties reproduce the measured experimental values \cite{Santos:2009yj}.\\

\paragraph*{Phenomenological Corrections}

Firstly, The amplitude is corrected for the real part as it is approximated to be purely imaginary. The real part of the amplitude is taken into account  by multiplying the cross section by a factor of (1+$\beta^2$) with $\beta = \tan\big(\lambda \pi/{2})$, and $\lambda = \partial \log (\mathcal{A}_{T,L}^{\gamma^*p\rightarrow Vp})/\partial \log(1/x_{I\!\!P})$. Secondly, the exchanged two gluons may have different momentum fractions, so a skewedness correction \cite{Shuvaev:1999ce} to the amplitude is applied, by a factor $R_g(\lambda)= 2^{2 \lambda +3}/\sqrt{\pi} \cdot\Gamma (\lambda_g + 5/2)/\Gamma (\lambda_g+4)$ with $\lambda_g= \partial \log (x_{I\!\!P}g(x_{I\!\!P}))/\partial \log(1/x_{I\!\!P})$.\\

As a test of the IPSat+QMC model against the only available measurement of the $|t|$ dependence of coherent $J/\psi$ photo production off a heavy nucleus, Fig.~\ref{comp} compares our calculation with the ALICE data taken in ultra-peripheral Pb--Pb collisions at $\sqrt{s_{NN}} = 5.02$~TeV at midrapidity, $-0.8 \le y \le 0.8$~\cite{ALICE:2023gcs}.
The data are well described within the quoted uncertainties.
The measured range, $|t| \le 0.012$~GeV$^2$, lies entirely inside the first diffractive cone of lead, below the first minimum of the coherent spectrum.

The comparison thus anchors the dipole dynamics and the average transverse size of the target in the relevant small-$x$ regime, while indicating that current UPC data are not sufficiently sensitive to distinguish between interaction-informed and phenomenological density profiles. This is expected for heavy nuclei, where shell-structure effects are comparatively weak. Residual sensitivity may nevertheless appear near the third or fourth diffractive minimum, although accessing this region experimentally is challenging because incoherent contributions become increasingly important at large $|t|$.

\begin{figure}[h]
    \centering
    \includegraphics[width=0.95\columnwidth]{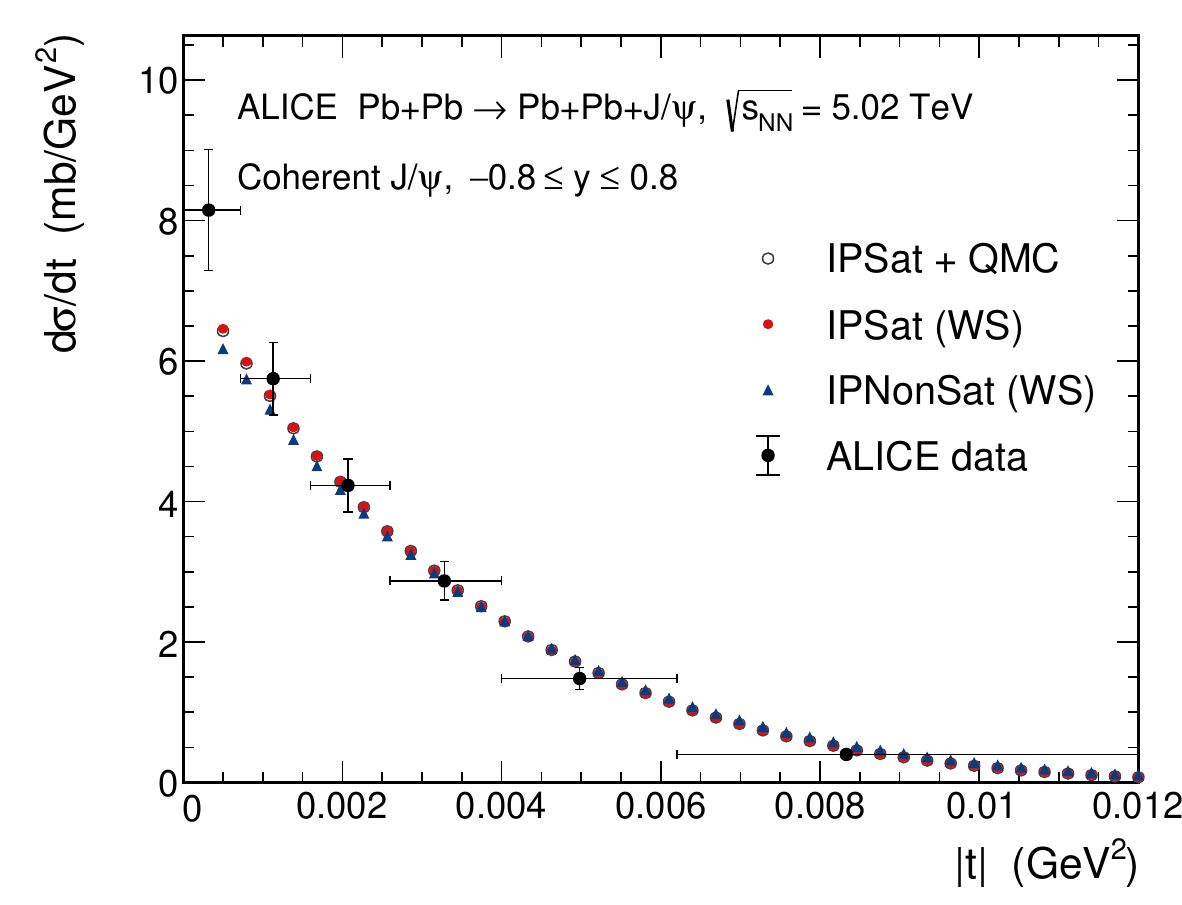}
    \caption{
    Differential cross section for coherent $J/\psi$
    photoproduction in Pb--Pb collisions as a function of the
    squared momentum transfer, compared with ALICE data~\cite{ALICE:2023gcs}.
    }
    \label{comp}
\end{figure}

\end{document}